# Harnessing Ferro-Valleytricity in Penta-Layer Rhombohedral Graphene for Memory and Compute

Md Mazharul Islam[1], Shamiul Alam[1], Md Rahatul Islam Udoy[1], Md Shafayat Hossain[2], Kathleen E Hamilton[3], and Ahmedullah Aziz[1*]

[1] Dept. of Electrical Eng. and Computer Sci., University of Tennessee, Knoxville, TN, 37996, USA
[2] Dept. of Physics, Princeton University, Princeton, NJ, 08544, USA
[3] Oak Ridge National Laboratory, Oak Ridge, TN 37831, USA
[*]Corresponding Author. Email: aziz@utk.edu

***Abstract*- Two-dimensional materials with multiple degrees of freedom, including spin, valleys, and orbitals, open up an exciting avenue for engineering multifunctional devices. Beyond spintronics, these degrees of freedom can lead to novel quantum effects such as valley-dependent Hall effects and orbital magnetism, which could revolutionize next-generation electronics. However, achieving independent control over valley polarization and orbital magnetism has been a challenge due to the need for large electric fields. A recent breakthrough involving penta-layer rhombohedral graphene has demonstrated the ability to individually manipulate anomalous Hall signals and orbital magnetic hysteresis, forming what is known as a valley-magnetic quartet. Here, we leverage the electrically tunable Ferro-valleytricity of penta-layer rhombohedral graphene to develop non-volatile memory and in-memory computation applications. We propose an architecture for a dense, scalable, and selector-less non-volatile memory array that harnesses the electrically tunable ferro-valleytricity. In our designed array architecture, non-destructive read and write operations are conducted by sensing the valley state through two different pairs of terminals, allowing for independent optimization of read/write peripheral circuits. The power consumption of our PRG-based array is remarkably low, with only ~ 6 nW required per write operation and ~ 2.3 nW per read operation per cell. This consumption is orders of magnitude lower than that of the majority of state-of-the-art cryogenic memories. Additionally, we engineer in-memory computation by implementing majority logic operations within our proposed non-volatile memory array without modifying the peripheral circuitry. Our framework presents a promising pathway toward achieving ultra-dense cryogenic memory and in-memory computation capabilities.**

## Introduction

The manipulation of the electron's spin degree of freedom has given rise to the field of spintronics which led the researchers to create and envision spin-based functional devices [1], [2]. In several solids, such as Si, Ge, Bi and in several 2D materials (AlAs, graphene, and $MoS_2$), electrons exhibit an additional property known as valley, where they occupy multiple minima in the conduction band or valleys in the Brillouin zone [3]–[8]. This discovery has paved the way for valleytronics, wherein the manipulation of electron valley occupancy enables the creation of functional devices [7]. Throughout the last two decades, researchers have been investigating the field of valleytronics to uncover its theoretical and experimental aspects to explore a wide range of practical applications [9], [10]. Valley polarization, due to its orbital nature, is more easily achieved through external stimuli such as electric fields, magnetic fields, and circular polarization of light [11]–[13]. Similarly, orbital magnetization, another characteristic of electrons, depends on quantum geometric effects and can be controlled in the same way as valley polarization [11]–[14]. However, realizing functional electronic devices based on valley polarization is challenging, typically requiring large electric fields for the valley switching process [15]–[19]. Researchers are constantly exploring for new device structures across various domains to achieve independent tunability of valley-dependent Hall effect and orbital magnetism [20], [21]. However, this endeavor remains challenging due to the intertwined nature of these phenomena and material constraints [22]–[26]. Han *et al.* have recently demonstrated a four-terminal device based on pentalayer rhombohedral graphene (PRG) that can offer independent tunability of valleys and orbital magnetization thereby forming a valley-magnetic quartet [27]. While this device holds promise for advancing research in valleytronics at cryogenic temperatures, its potential for functional device applications, particularly in memory and logic operations, remains unexplored [27].

Given that valley tuning experiments are typically demonstrated at low temperatures, it is natural to explore cryogenic applications of valleytronic phenomena. In fact, cryogenic devices have garnered significant attention in recent years due to their remarkable power efficiency and ultrafast switching characteristics [28]–[32]. Cryogenic memory technologies, capable of functioning at temperatures below 4 Kelvin (K), are widely being researched in both classical and quantum computing frameworks[33], [34]. Leveraging the PRG-based device for cryogenic operation can catalyze the research prospects of cryogenic valleytronics for next-generation memory and logic devices with superior performance. Here, we exploit the non-volatile valley state of the PRG device to propose ultra-low power, selector-less non-volatile memory device for cryogenic application. Additionally, we engineer in-memory computation within our proposed PRG-based array architecture by implementing in-memory majority logic operations.

## Device Characteristics and Modeling Approach of Pentalayer Rhombohedral Graphene

To investigate orbital multiferroicity, the device utilizes stacked three layers of PRG sheets interleaved with hexagonal boron nitride (hBN) layers (Fig. 1(a)) [27]. This material configuration is essential for fabricating a dual-gated Hall bar device (Fig. 1(a)), enabling independent manipulation of valley and orbital magnetization, thereby forming a valley-magnetic quartet. By introducing hole-doping into the graphene layer, orbital magnetic hysteresis was successfully induced under a constant gate electric field. Furthermore, sweeping the gate electric field triggers hysteretic valley polarization, leading to bipolar hysteretic Hall voltages. We leverage the characteristics of valley-driven hysteretic Hall voltage in our proposal, while reserving the utilization of orbital magnetization switching for future research endeavors.

For our investigation, we have developed a Verilog-A-based behavioral circuit model of the PRG-based Hall bar device that's compatible with SPICE, enabling circuit-based simulation-driven analysis. The model effectively captures the behavioral dependency of the Hall resistance on the applied gate electric field as reported in [27] (Figs. 1(b,c)). Fig. 1(c) depicts the characteristic of $R_{xy}$ as a function of the applied electric field showcasing the behavioral accuracy of our model. We identify two critical electric fields ($E_{C+}$ and $E_{C-}$) necessary for valley polarization switching (Figs. 1(b,c)). To accurately predict the induced Hall voltage ($V_{xy}$) (Figs. 2(a,c)), we have extracted the longitudinal resistance ($R_{xx}$) for zero applied magnetic field as shown in Fig. 2(b). For conducting circuit-level simulations of our proposed memory and logic circuit, we utilize HSPICE, a cutting-edge circuit simulation tool [35]. Further elaboration on the operation of the proposed memory, logic, and logic-in-memory is provided in the subsequent sections.

## PRG-based Non-volatile Memory Array

We start our discussion by delineating the dynamics of PRG-based dual gated Hall bar device and how it can be utilized for non-volatile memory array application. The structural arrangement of the device and our proposed array structure is depicted in Figs. 3(a,b), while Fig. 3(c) illustrates the symbol adopted for illustrative representation. When an electric field is applied between the two gates, it induces a longitudinal current across the PRG layer, leading to a Hall voltage between the transverse terminals of the device (Fig. 2(c)). Fig. 1(b) depicts the electric field dependency of the valley polarization in the device under consideration. We define the electron valley state K and K' as ‘0’ and ‘1’ respectively. Based on this definition, we can delineate three distinct regions of applied gate voltage ($V_{GATE}$) (Fig. 2(c)). (i) $V_{GATE} \leq V_{C-}$: write ‘0’ region, (ii) $V_{C+} \leq V_{GATE} \leq V_{C+}$: read region, and (iii) $V_{GATE} \geq V_{C+}$: write ‘1’ region. At a critical switching electric field ($V_{C+}$) of ± 16 mV, a transition occurs in the valley polarization, switching between K and K' valleys. We exploit this hysteretic valley polarization to propose a novel non-volatile memory structure. Here, the write operation of ‘0’ and ‘1’ requires opposite polarity of $V_{GATE}$, whereas for the read operation, we only use positive voltage for the read operation ($V_{C+} \leq V_{READ} \leq V_{C+}$). The trajectory of the Hall voltage variation with the applied electric field for both the valley polarizations are illustrated in Fig. 2(c). As shown in the figure, two different signs of Hall resistance ($R_{xy}$) are observed for the two different valley polarizations. The induced Hall voltage can be estimated as, $V_{HALL} = \frac{V_{READ}}{R_{xx}} \times R_{xy}$, where $R_{xx}$ is the longitudinal resistance of the PRG memory cell (Fig. 2(b)). Consequently, two different signs of Hall voltages across the memory cell's transverse terminals are induced for the two distinct stored memory states (Fig. 2(c)). Fig. 2(a) presents the simulation waveform illustrating the read and write operations of a single PRG-based memory device.

To evaluate the practicality of a memory, it is crucial to design an array architecture and investigate its array-level performance. Fig. 3(b) illustrates our proposed memory array architecture. Each memory cell is sandwiched between orthogonally running bit-lines (BLs) and word-lines (WLs). The transverse terminals of each memory cell are cascaded to obtain the accumulated induced Hall voltage in a single row (Fig. 3(b)). A single row-output is fed to a comparator input to amplify the accumulated Hall voltages in a single row. Each word line is linked to the top gates (TG) of the PRG cells, while the bottom gates (BG) of each column are connected to individual bit-lines (BL). The DC voltage difference between the TG and BG effectively dictates the applied electric field inducing a transverse Hall voltage that is subsequently sensed in the output comparator. Our proposed array architecture eliminates the need for selector devices for individual memory cell access. We introduce distinct biasing schemes for both read and write operations, facilitating a selector-less memory array design. Selector-less memory devices offer potential for highly scalable memory technology with reduced cost and complexity. During read operations, our array architecture accesses each column independently, while during write operations, it allows for individual access for each memory cell.

## Read/Write Operations

Figs. 3(d,e) provide a comprehensive overview of the complete read operation in our proposed array architecture. As shown in Fig. 2(c), non-zero Hall voltage is induced for a non-zero applied DC voltage. Hence, it is crucial to ensure that all other cells in a single row, except the one being accessed, receive zero electric field or zero DC voltage difference. In our proposed biasing scheme for the read operation, all the unaccessed cells in a single row have zero DC voltage difference between the TG and BG. To read

the stored data from the accessed column, an access voltage ($V_{ACC}$) is applied to all the rows (WLs) (Figs. 3(d,e)). In our analysis, we assume the first column as the accessed column and analyze its read operation. All the unaccessed columns receive a non-zero read voltage, $V_{ACC} = V_{READ}$. This way, all the cells in the accessed column receive a DC voltage difference of $V_{READ}$ = 10 mV, and the corresponding Hall voltage of each of these cells are reflected in the comparator input after amplification. If the *k*th cell of the *i*th row is being accessed for the read operation, the resultant terminal voltage at the amplifier input can be expressed as:

$$V_{hr-i} = \sum_{j=1}^{n} V_{xy-ij} - 2n \times V_{loss} = \frac{V_{bias-ik}}{R_{xx-ik}} \times R_{xy-ik} - 2n \times V_{loss}$$

Here, *n* is the total number of PRG cells in a single row and $V_{loss}$ is the voltage drop in the Cr/Au connectors. The induced Hall resistance in all the unaccessed cells within a single row is effectively zero due to the application of zero longitudinal DC voltage across them. In our proposed array architecture, $V_{READ}$ is set to 8 mV, ensuring reliable read operation with maximum read margin of ~8 mV. In our analysis, we designate the first column as the accessed column. Fig. 3(e) illustrates the read operation of the first column. Our proposed PRG-based memory array consumes ~2.3 nW per read operation per cell which is significantly lower than that of the state-of-the-art cryogenic memories[36]–[40] All unaccessed cells in our designed array architecture receive zero voltage, effectively addressing concerns related to leakage signals commonly encountered in partially accessed (i.e., half-accessed) cells in conventional memory devices.

Figs. 3(d,f) offer a brief overview of the write scheme implemented in our proposed array architecture. For individual memory cell access during write operations, we employ a standard biasing scheme known as V/2 biasing. In this setup, the WL of the targeted cell receives a write voltage $V_{WRITE}$, such that $V_{WRITE} \geq V_{C+}$ for 0→1 transition and $V_{WRITE} \leq V_{C-}$ for 1→0 transition. All the other WLs are kept at $\frac{V_{WRITE}}{2}$. To ensure that the targeted cell receives a full switching voltage, the corresponding BL is set to 0V, while all the other BLs receive a voltage of $\frac{V_{WRITE}}{2}$. The other memory cells in the accessed row or column are referred to as half-accessed cells, as the inactive cells in this row or column receive half of the access voltage. All the other cells ideally have zero voltage across them. As an example, we consider $M_{1,1}$ as the accessed cell to examine its write operation for 0→1 transition as illustrated in Fig. 3(b). Here, only the accessed cell ($M_{1,1}$) exhibits a transition of the valley state which is reflected by the change of the Hall voltage sign during read operation. All the half-accessed and unaccessed cells remain unaffected. Here, $V_{READ}$ has the same sign as $V_{WRITE}$ required for 0→1 transition. Hence, the chosen polarity of $V_{READ}$ prevents the accidental 1→0 data flip. Moreover, the magnitude of the read voltage is about half of the critical voltage required for transition ($V_{C+}$), alleviating the possibility of the accidental data overwrite of 0→1. The power consumption for the write operation is estimated to be ~6 nW per cell, which is orders of magnitude lower than the state-of-the-art cryogenic memory technologies[36], [37].

**In-memory Computation based on PRG-based Non-volatile Memory Array**

Cryogenic in-memory computation offers substantial advantages in terms of energy efficiency and speed due to the characteristics of the cryogenic environment. Additionally, in-memory computation serves as a promising solution to overcome the bottlenecks associated with von-Neumann architecture and the memory wall. In light of the ongoing trend of exploring non-volatile devices for in-memory computation, here, we explore cryogenic in-memory Majority logic based on the PRG-based Hall bar device. Majority logic is a Boolean logic primitive which generates an output of logic '1' when there is a greater number of logic '1' in the inputs. It is a fundamental component for implementing various functions such as adders, multiplexers, encoders, and decoders. Majority function along with the NOT gate creates a functionally complete set, and, in several technologies, majority logic is easier to implement. Thanks to these advantages, Majority logic has been extensively explored for in-memory computational architectures.

In Fig. 4(a), we present the schematics of a 5-input Majority gate implemented within a single row of the previously discussed array setup. The comparator reference is set to 0V to sense the polarity of the induced Hall voltage at the row output as shown in Fig. 4(a,b). The simulation results for all possible input combinations are illustrated in Fig. 4(b), showcasing six different levels of voltages at the output of the voltage amplifier after applying suitable $V_{Bias}$ to the input storage cells. We leverage the bipolarity of the PRG Hall resistance to create a linear relationship between the array output voltage and number of 'logic 1' in a single row (Fig. 4(c). When we apply the sense enable signal ($V_{EN}$) to the sense amplifiers, we obtain the output of a 5-input majority gate at the sense amplifier output illustrated in Figs. 4(b,c). Similar Majority operations with varying numbers of inputs can be designed by energizing the corresponding input storage cells in a single row. Fig. 4(d) summarizes the output Hall voltages and the corresponding output logic value for 3-input and 5-input majority operation in our PRG-based Hall bar cells. Our proposed majority operation provides a separation of about ~45 $m$A for sensing purposes.

## Discussion and Conclusion

The distinctive ferro-valleytricity characteristic offers an avenue for developing innovative functional devices with remarkable attributes. Our work leverages this feature in a device based on 2D pentalayer rhombohedral graphene, specifically tailored for non-volatile memory applications in cryogenic environments. The Hall resistance was measured in a Bluefors LD250 dilution refrigerator at 100 mK as reported in [27]ss. We have developed a behavioral circuit model that accurately captures the non-volatile valley state and estimates the induced transverse Hall voltage in response to an applied DC bias. The device employs two separate pairs of terminals for its read and write operation. Leveraging this unique feature, we propose and analyze a unique memory array architecture. Our proposed biasing schemes show successful column-wise read operations and individual write operations. Additional analog components required for complete read and write operations are integrated into the array peripherals. Furthermore, we have engineered in-memory Majority logic operations using our proposed memory array. The architecture allows for the implementation of in-memory Majority functionality without necessitating modifications to the peripheral circuitry. Our proposed non-volatile memory architecture stands out as a promising choice for cryogenic memory systems due to its low operating temperatures. Notably, cryogenic memory block serves as a pivotal element in quantum computing setups and space electronics, where our proposed device can make a substantial impact.

## Data Availability

The data that support the plots within this paper and other finding of this study are available from the corresponding author upon reasonable request.

## Author Contributions

M.M.I conceived the idea, designed the memory array, and performed the simulations. S.A., M.R.I.U., analyzed and helped finalizing the designs. M.S.H. provided insights into the majority logic operation. A.A. supervised the project. All authors commented on the results and wrote the manuscript.

## Competing Interests

The authors declare no competing interests.

## Acknowledgement

The work of Md Mazharul Islam and Shamiul Alam was supported by the Science Alliance, a Tennessee Higher Education Commission Center of excellence administered by The University of Tennessee-Oak Ridge Innovation Institute on behalf of The University of Tennessee Knoxville.

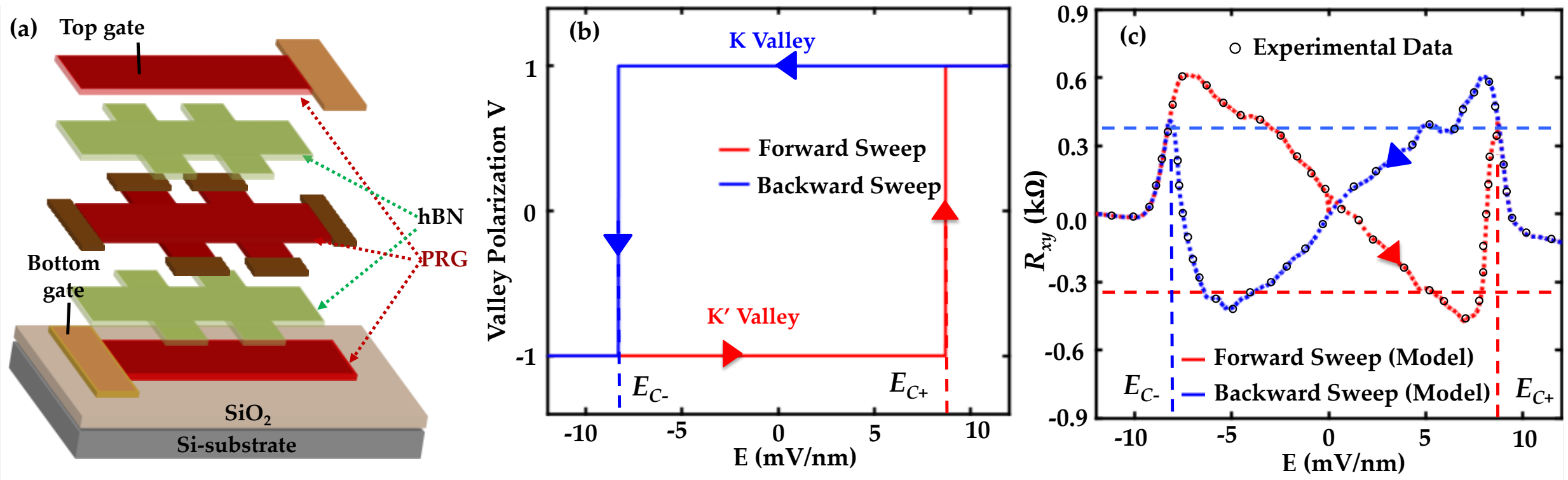

**Fig. 1. Device characteristics of the pentalayer rhombohedral graphene (PRG)-based Hall bar device. (a)** Schematics of the dual-gated device. **(b)** Valley polarization plot illustrating the dependence on the applied electric field, with marked switching values of the electric fields ($E_{C+}$and $E_{C-}$). **(c)** Transverse Hall resistance ($R_{xy}$) characteristics induced by the applied gate electric field. The solid curve represents the characteristics of the developed behavioral model ('red' curve for forward sweep and 'blue' curve for backward sweep), and dotted values indicate the experimental data. The measurement was measured in a dilution refrigerator with an electronic temperature of around 100 mK [30].

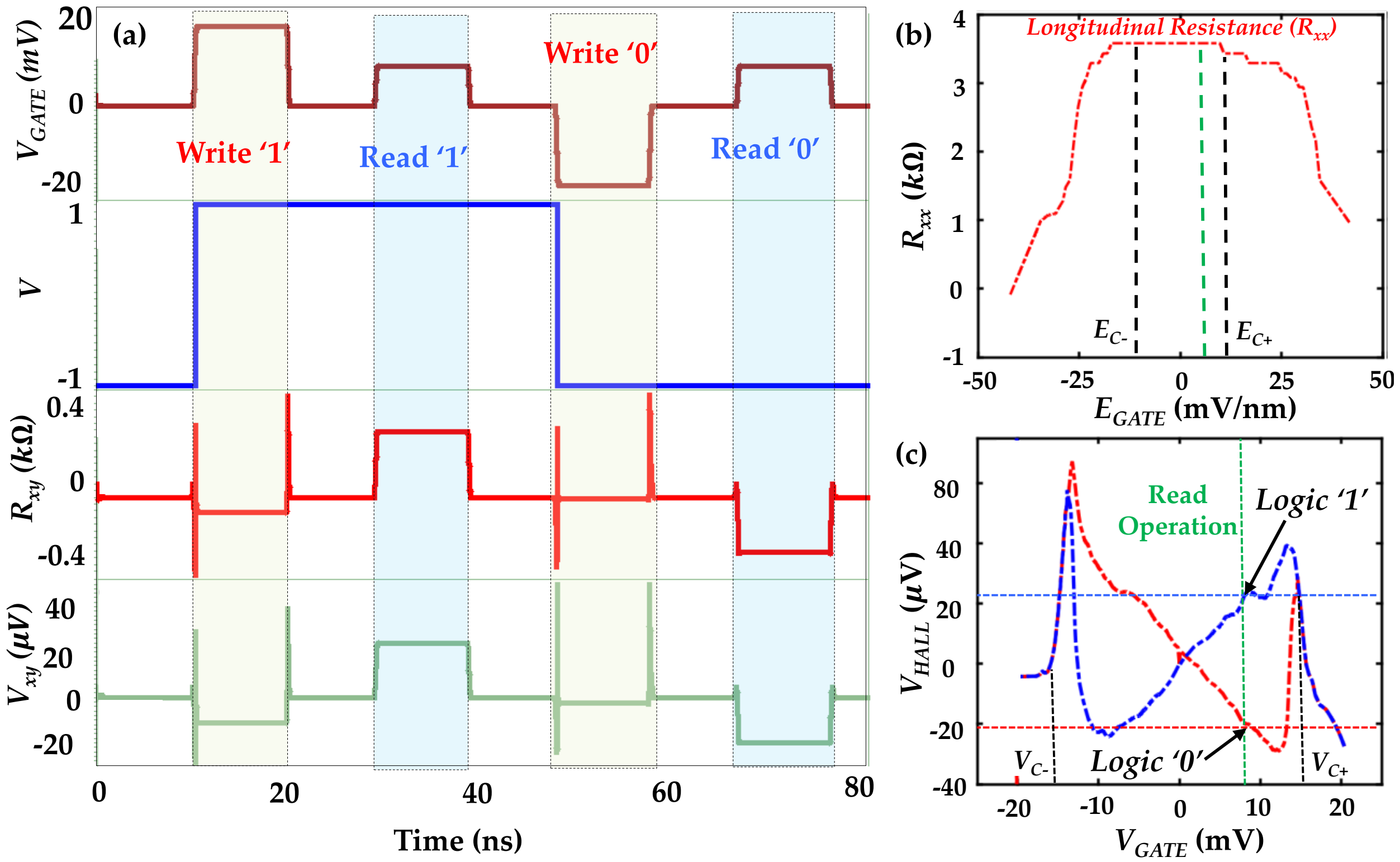

**Fig. 2. Memory operation of the PRG-based Hall bar device. (a)** Simulation waveform illustrating the read/write operation of an individual PRG-based Hall bar device. The write '1' operation is executed followed by a read operation. Subsequently, a write '0' operation is conducted, again followed by a read operation, demonstrating the induction of two different polarities of Hall voltages for distinct logic states. **(b)** Extracted longitudinal resistance ($R_{xx}$) utilized for estimating the Hall voltage ($V_{HALL}$) depicted in **(c). (c)** Hall voltage ($V_{HALL}$) induced across the transverse terminals of the PRG-based Hall bar device during the read operation. The applied DC bias chosen for the read operation ($V_{READ}$) is marked.

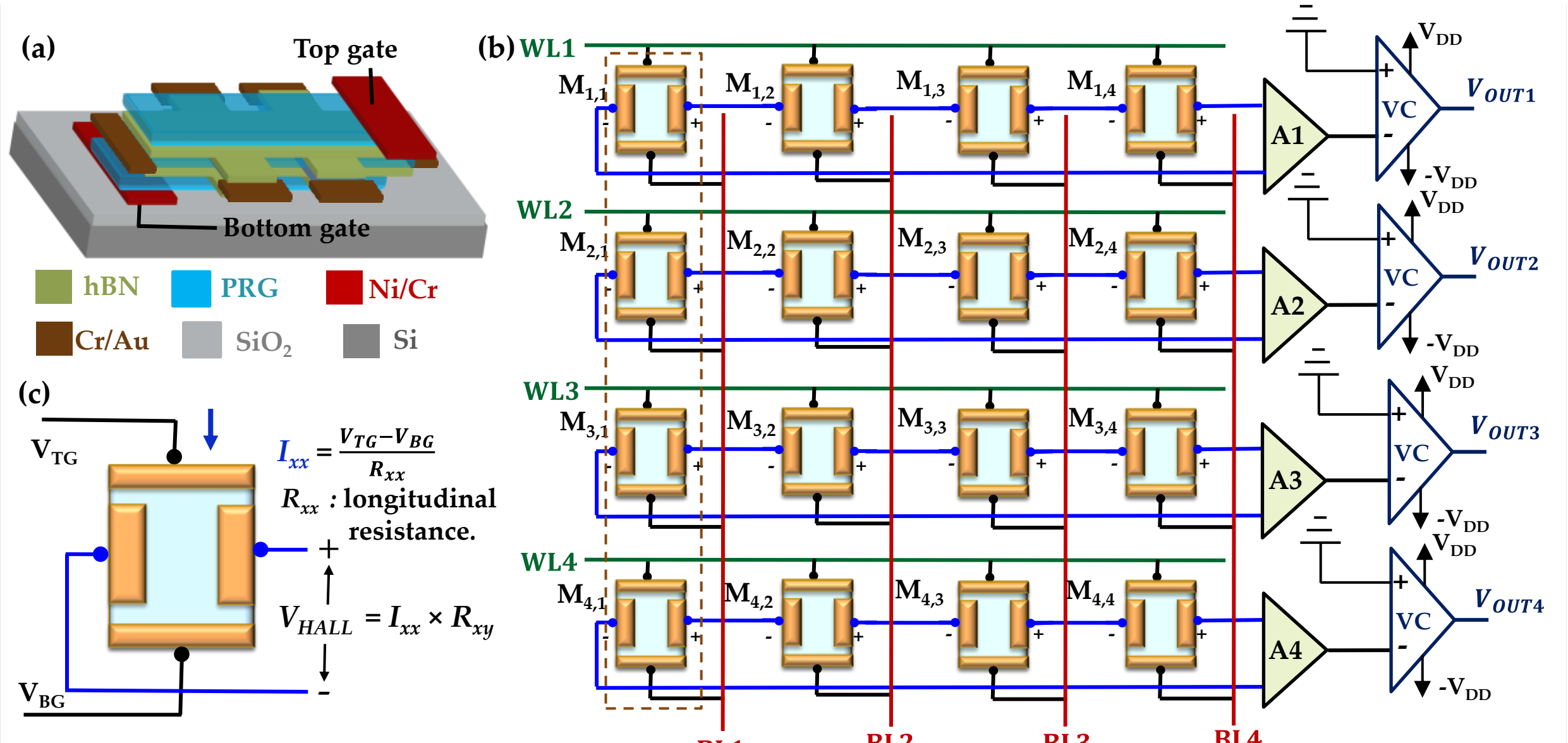

**(d) Table I:** Biasing Scheme for read/write operation

| Operation | | $V_{WL}$ | $V_{BL}$ | $\Delta V_{xx}$ |
|---|---|---|---|---|
| Write | 0 → 1 | $V_{WRITE}$: accessed row<br>$\frac{V_{WRITE}}{2}$ : un-accessed row | 0 : accessed column<br>$\frac{V_{WRITE}}{2}$ : un-accessed column | $V_{WRITE}$ : accessed<br>$\frac{V_{WRITE}}{2}$ : half-accessed.<br>0 : unaccessed. |
| | 1 → 0 | $-V_{WRITE}$: accessed row<br>$\frac{-V_{WRITE}}{2}$ : un-accessed row | 0 : accessed column<br>$\frac{-V_{WRITE}}{2}$ : un-accessed row | $-V_{WRITE}$ : acc.<br>$\frac{-V_{WRITE}}{2}$ : half-acc.<br>0: unaccessed. |
| Read | | $+V_{READ}$ : All column | 0 : accessed column<br>$+V_{READ}$: un-accessed row | $+V_{READ}$ : acc. Column<br>0: unacc. Column |

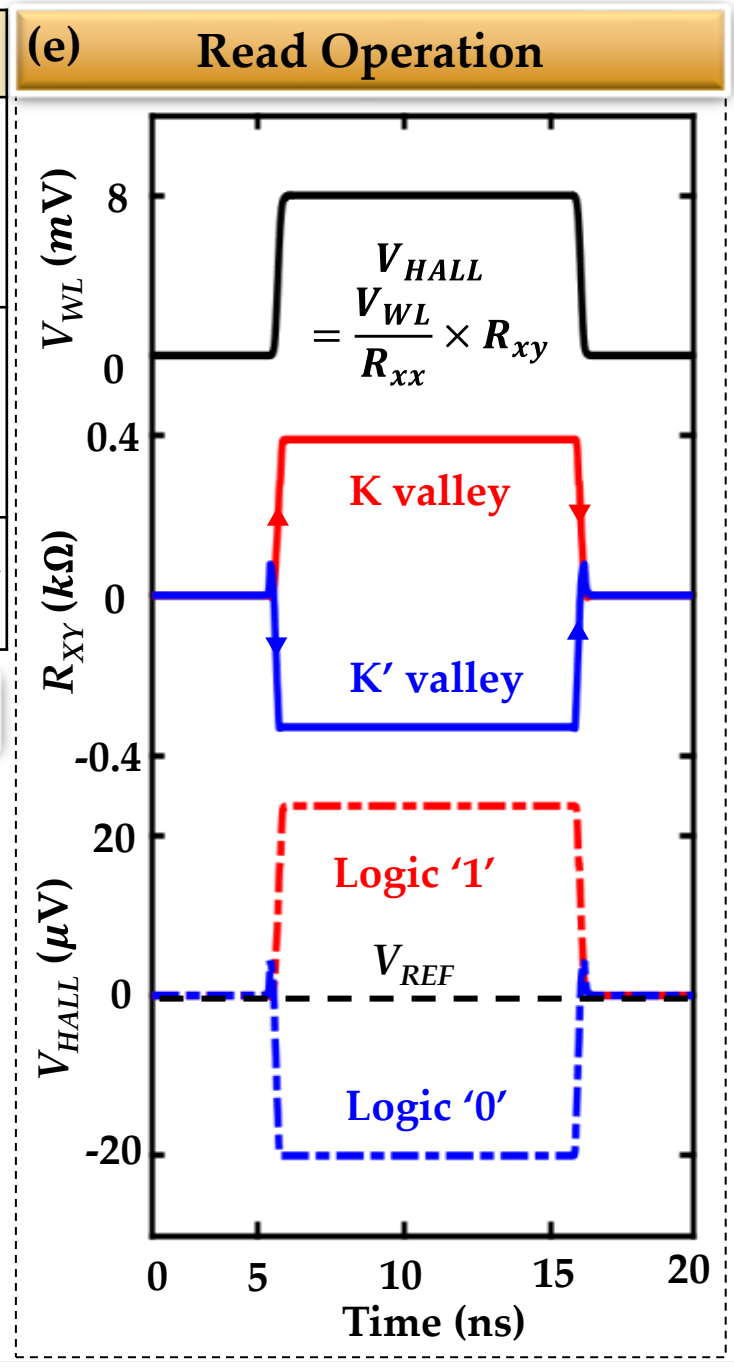

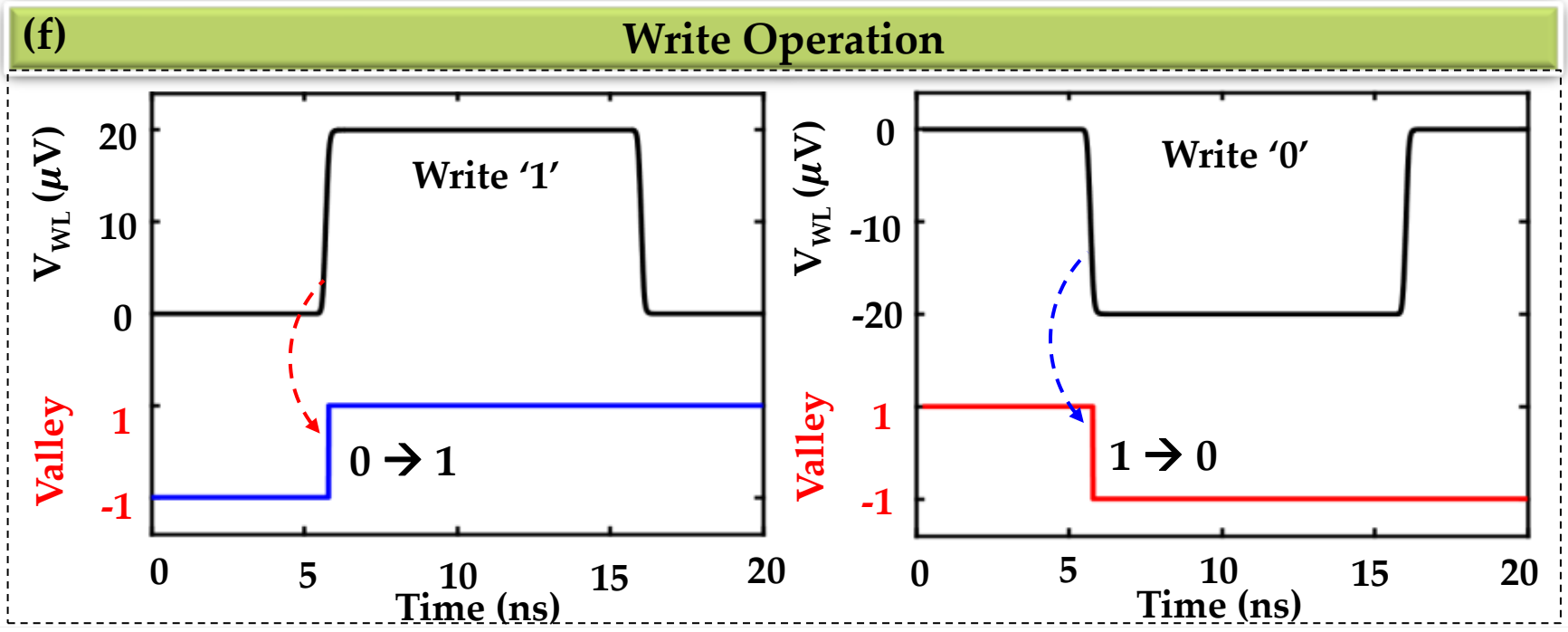

**Fig. 3. Array-level operation of the PRG-based memory. (a)** Schematics of the PRG-based Hall bar device at the device level. **(b)** Proposed array architecture based on dual-gated Hall bar device. The four terminal device is positioned between perpendicular word lines (WLs) and bit lines (BLs). The transverse Hall terminals are cascaded to accumulate the total $V_{HALL}$ in a single row. **(c)** Symbol representation of the dual-gated PRG-based four-terminal Hall bar device. **(d)** Summary table (Table I) of biasing schemes for read/write operations in the depicted array architecture depicted in **(b)**. **(e)** Simulation waveform showing the read operation of the first column. **(f)** Simulation waveform illustrating the write operation of the accessed $M_{1,1}$ cell. Here, the valley polarization is illustrated instead of the induced Hall resistance.

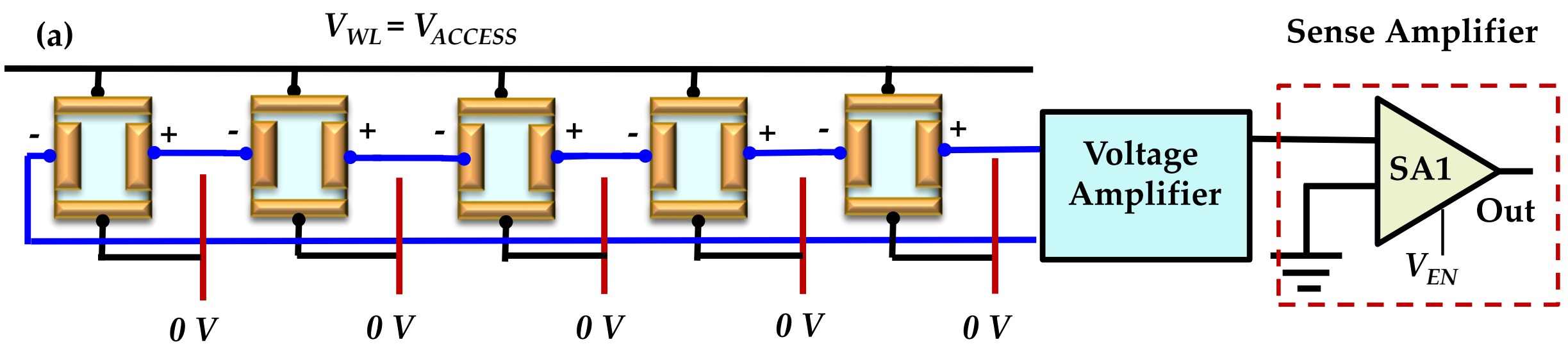

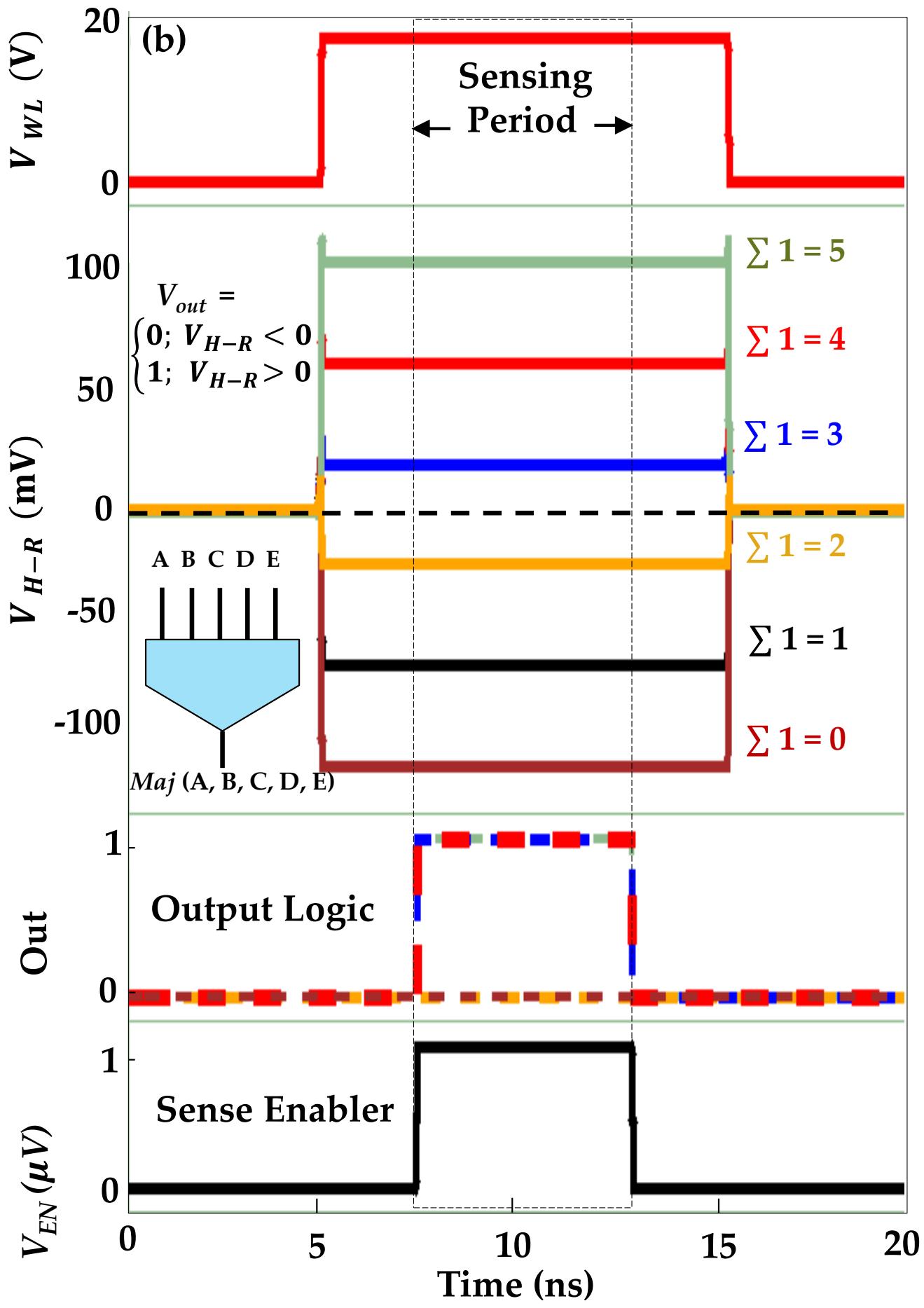

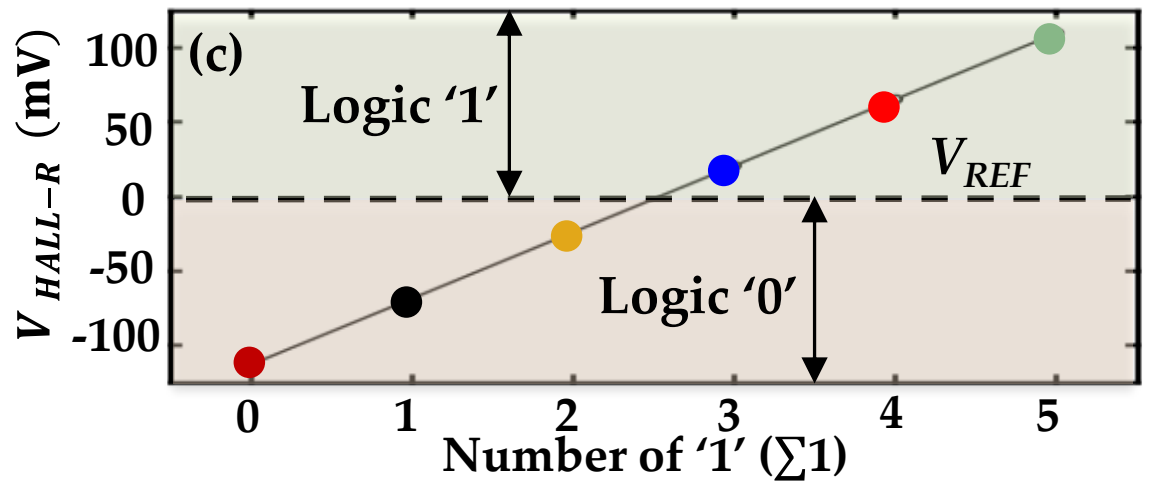

**(d) Table II:** Summary of Majority Gate Operation.

| # Inputs | ∑ inputs | ∑ $V_{HALL}$ | Output Logic |
|---|---|---|---|
| 3 input | 0 | - 68.6 mV | 0 |
| | 1 | -23.8 mV | |
| | 2 | +20.8 mV | 1 |
| | 3 | +65.5 mV | |
| 5 input | 0 | -113 mV | 0 |
| | 1 | -68.6 mV | |
| | 2 | -23.9 mV | |
| | 3 | +20.8 mV | 1 |
| | 4 | +65.5 mV | |
| | 5 | +110 mV | |

**Fig. 4. Majority gate operation utilizing the PRG-based memory array. (a)** Schematics of a single row storing inputs for the 5-input majority gate operation. **(b)** Simulation waveform of a 5-input majority carrier operation for all possible input combinations. The output of the sense amplifier becomes '0' or '1' depending on the polarity of the accumulated Hall voltage ($V_{HALL-R}$). In our current model, we have not included the switching delay due to insufficient information. However, integrating the switching delay into our model in the future is a feasible endeavor. **(c)** Different levels of accumulated induced Hall voltage ($V_{HALL-R}$) corresponding to different numbers of 'logic 1' inputs at the input terminal, demonstrating a linear relationship. **(d)** Summary table (Table II) detailing the output $V_{HALL-R}$ and corresponding output logic for 3-input and 5-input majority gates.